\documentclass[11pt,a4paper]{article}
\usepackage[T1]{fontenc}
\usepackage{lmodern}
\usepackage[margin=28mm]{geometry}
\usepackage{amsmath,amssymb,amsthm,mathtools}
\usepackage{booktabs,array,tabularx,longtable}
\usepackage{graphicx,xcolor}
\usepackage{enumitem}
\usepackage[numbers,sort&compress]{natbib}

\usepackage{pgfplots}
\pgfplotsset{compat=1.18}
\usepackage{placeins}
\usepackage{flafter}
\usepackage[colorlinks=true,allcolors=blue!45!black]{hyperref}
\usepackage[capitalise,noabbrev]{cleveref}
\newtheorem{theorem}{Theorem}[section]
\newtheorem{proposition}[theorem]{Proposition}

\newtheorem{assumption}[theorem]{Assumption}
\theoremstyle{definition}

\theoremstyle{remark}

\newcommand{\Var}{\operatorname{Var}}

\title{Residual Curvature and Wald Calibration in\\Autocorrelation-Based Timescale Estimation}
\author{Yongsik Sim\thanks{Correspondence: \href{mailto:y.sim@skku.edu}{y.sim@skku.edu}. ORCID: \href{https://orcid.org/0000-0003-2711-2793}{0000-0003-2711-2793}.}\\[0.4em]
\small Department of Radiology and Center for Imaging Sciences,\\
\small Samsung Medical Center, Sungkyunkwan University School of Medicine,\\
\small Seoul, Republic of Korea}

\date{16 November 2026}
\begin{document}
\maketitle
\begin{abstract}
Autocorrelation-based timescale estimation often fits a single exponential to a finite vector of sample autocorrelations, even when the underlying process is not autoregressive of order one. We examine the first-order uncertainty of this fixed-lag projection. Under misspecification, its sandwich denominator is the full population curvature, including a residual--second-derivative term; a Gauss--Newton denominator need not agree with it. We apply standard extremum-estimation arguments to the autocorrelation normalization, establish a conditional sign rule, and give two stationary Gaussian examples with exact standard-error ratios $14/13$ and $13/15$. The discrepancy can therefore produce either underestimation or overestimation of uncertainty. The full-to-Gauss--Newton population standard-error ratios are 1.34 and 1.45 in the two $K=36$ noise-mixture models, and range from 0.38 to 1.37 across the evaluated published-process parameter grid. In a follow-up experiment with 24,000 synthetic series, empirical standard deviations in four Gaussian models are consistent with the full-curvature scale at the main sample length, whereas the Gauss--Newton scale is substantially too small in the two noise-mixture models. In all 4,000 observed replicates of the primary condition, full-curvature intervals contained their Gauss--Newton counterparts, constraining the observed coverage gain to be nonnegative. Fixed-bandwidth HAC estimation converges to a smoothed long-run variance, and using the full sample curvature can worsen finite-sample coverage in an exactly specified model. Bandwidth-specific results, directional coverage errors, and oracle comparisons distinguish curvature effects from long-run variance estimation and transformation effects. These findings identify an application-specific discrepancy in a published autocorrelation-domain variance formula and characterize its asymptotic and finite-sample consequences.
\end{abstract}

\section{Introduction}\label{sec:introduction}
Fitting an exponential curve to an autocorrelation function provides a low-dimensional description of temporal dependence. When the observed autocorrelation is not exactly exponential, the fitted parameter describes a projection of the autocorrelation sequence. Uncertainty quantification must therefore refer to this projection target, which may differ from a latent autoregressive coefficient. This distinction is relevant to autocorrelation-domain (AD) methods for estimating timescales from functional magnetic resonance imaging.

\citet{Riegner2026} define a fixed-$K$ exponential-ACF projection and propose a sandwich variance with a Gauss--Newton denominator for inference under model misspecification. We examine the first-order variance of this AD estimator and the finite-sample behavior of the associated Wald intervals. The time-domain (TD) linear projection estimates a different target and has a different estimating-equation sensitivity.

Misspecification-robust inference based on the full criterion Hessian is established for independent, identically distributed observations \citep[Theorem~3.3]{White1981}. The asymptotic theory of nonlinear least squares with dependent observations is also established \citep{WhiteDomowitz1984}. Joint asymptotic inference for sample autocorrelations under weak dependence is classical \citep{RomanoThombs1996}. For the ACF projection considered here, the derivation combines a joint sample-moment limit with the effect of ACF normalization. The resulting score variance is a long-run variance, commonly estimated by a heteroskedasticity- and autocorrelation-consistent (HAC) procedure \citep{NeweyWest1987}. Curvature and long-run variance estimation are distinct determinants of interval calibration.

Applying standard implicit-function and delta-method arguments, we obtain the fixed-$K$ influence expansion and its full curvature. We establish a conditional rule for the direction of the Gauss--Newton discrepancy and construct stationary Gaussian counterexamples in both directions. A simulation study compares empirical sampling variability with the population first-order scales and examines feasible Wald coverage across bandwidths. Fixed-bandwidth limits, oracle substitutions, and directional miss probabilities distinguish the denominator discrepancy from other sources of miscalibration.

\section{Fixed-lag autocorrelation projection}
\label{sec:theory}
Let $(X_t)_{t\in\mathbb Z}$ be stationary with mean $\mu$. Write
$Y_t=X_t-\mu$, $\gamma_k=\mathbb E(Y_tY_{t-k})$, and
$\rho_k=\gamma_k/\gamma_0$, where $\gamma_0>0$. For a fixed integer $K\geq1$,
\begin{equation}
 S_K(p)=\sum_{k=1}^K\{\rho_k-m_k(p)\}^2,\quad m_k(p)=p^k,\quad
 p^\star=\operatorname*{arg\,min}_{p\in\mathcal P}S_K(p),
 \label{eq:population-projection}
\end{equation}
where $\mathcal P$ is a compact interval in $(-1,1)$. This is a projection
of the first $K$ autocorrelations, not necessarily a generating autoregressive
coefficient. We omit the factor $1/K$ in the published objective
\cite{Riegner2026}; a common scaling of the objective cancels from a
consistently normalized sandwich variance.

With $\bar X_T=T^{-1}\sum_{t=1}^T X_t$, the biased sample ACF and estimator are
\begin{equation}
 \widehat\rho_k=
 \frac{\sum_{t=k+1}^T(X_t-\bar X_T)(X_{t-k}-\bar X_T)}
 {\sum_{t=1}^T(X_t-\bar X_T)^2},\qquad
 \widehat p=\operatorname*{arg\,min}_{p\in\mathcal P}
 \sum_{k=1}^K(\widehat\rho_k-p^k)^2.
 \label{eq:sample-projection}
\end{equation}
All lags share the variance denominator. Standardizing the centered series
by its sample standard deviation does not change this ACF.

\subsection{Regularity and exact curvature}
The asymptotic analysis uses the following joint moment conditions.
\begin{assumption}[Fixed-lag regularity]
\label{ass:fixed-k}
The process is strictly stationary and ergodic,
$\mathbb E|Y_0|^4<\infty$, $\gamma_0>0$, and
$\sqrt T\,\bar Y_T=O_p(1)$. For
\[
 V_t=(Y_t^2,Y_tY_{t-1},\ldots,Y_tY_{t-K})^\top,\qquad v=\mathbb E V_t,
\]
the entries of the covariance series are absolutely summable and
\begin{equation}
 T^{-1/2}\sum_{t=1}^T(V_t-v)\xrightarrow{d}N(0,\Sigma_V),
 \qquad \Sigma_V=\sum_{h\in\mathbb Z}\operatorname{Cov}(V_t,V_{t-h}).
 \label{eq:moment-clt}
\end{equation}
The minimizer $p^\star$ is unique and interior to $\mathcal P$.
The quantities $A$ and $\Omega$ below satisfy $A>0$ and $0<\Omega<\infty$.
\end{assumption}
The Gaussian autoregressive-plus-noise and moving-average examples below satisfy these conditions. Throughout the analysis, $K$ is fixed.

Set $R_k=\rho_k-m_k(p^\star)$ and $b_k=m_k'(p^\star)$. Then
\begin{align}
 G(p;\rho)&=\sum_{k=1}^K m_k'(p)\{\rho_k-m_k(p)\},
 &G(p^\star;\rho)&=0,\label{eq:score-equation}\\
 A&=-\partial_pG(p^\star;\rho)
 =\sum_{k=1}^K\{b_k^2-R_km_k''(p^\star)\},
 &q&=\sum_{k=1}^K b_k^2.\label{eq:bread}
\end{align}
Here $S_K''(p^\star)=2A$; $q$ is the Gauss--Newton curvature.
Derivatives are polynomial derivatives, so $m_1''(p)=0$, including at $p=0$.
The sample counterparts are
\begin{equation}
 \widehat A=\sum_{k=1}^K\left[
 \{m_k'(\widehat p)\}^2-
 \{\widehat\rho_k-m_k(\widehat p)\}m_k''(\widehat p)\right],
 \qquad \widehat q=\sum_{k=1}^K\{m_k'(\widehat p)\}^2.
 \label{eq:sample-bread}
\end{equation}
The residual-curvature term pairs the residual with $m''$. The first-order
condition pairs it with $m'$, a different vector.

The following result follows from the joint sample-moment central limit theorem by the ratio delta method and an implicit-function expansion of $G(p;\rho)=0$.

\begin{theorem}[Delta-method limit for fixed-lag ACF projection]
\label{thm:asymptotic}
Under Assumption~\ref{ass:fixed-k}, $\widehat p\to_p p^\star$ and
\begin{equation}
 \sqrt T(\widehat p-p^\star)
 =\frac{1}{A\sqrt T}\sum_{t=1}^T u_t+o_p(1)
 \xrightarrow{d}N\!\left(0,\frac{\Omega}{A^2}\right),
 \label{eq:asymptotic}
\end{equation}
where
\begin{align}
 u_t&=\frac1{\gamma_0}\sum_{k=1}^K b_k
 \{Y_tY_{t-k}-(p^\star)^kY_t^2\},\label{eq:influence-score}\\
 \Omega&=\sum_{h\in\mathbb Z}\operatorname{Cov}(u_t,u_{t-h}).
 \label{eq:lrv}
\end{align}
Moreover, $\widehat A\to_p A$ and $\widehat q\to_p q$.
\end{theorem}
\begin{proof}
The finite-dimensional law of large numbers and compactness give uniform
convergence of the objective and consistency. The ACF ratio expansion,
including centering and fixed-lag boundary terms, gives
\[
 \sqrt T(\widehat\rho_k-\rho_k)
 =\frac1{\gamma_0\sqrt T}\sum_{t=1}^T
 (Y_tY_{t-k}-\rho_kY_t^2)+o_p(1).
\]
Details are in Appendix~\ref{app:linearization}.
Expanding $G(\widehat p;\widehat\rho)=0$ at $(p^\star,\rho)$ gives
$\sqrt T(\widehat p-p^\star)
 =A^{-1}\sum_k b_k\sqrt T(\widehat\rho_k-\rho_k)+o_p(1)$.
The replacement of $\rho_k$ by $(p^\star)^k$ in the weighted score is exact
because $\sum_k b_kR_k=0$. Equation~\eqref{eq:moment-clt} gives the limit,
and continuity gives the two curvature limits.
\end{proof}

The variance has the full-Hessian sandwich form of misspecified extremum estimation \cite[Theorem~3.3]{White1981}. For this ACF projection, the preceding expansion accounts for normalization and serial dependence through the joint moment central limit theorem. With the same consistent long-run variance estimator,
\begin{equation}
 \frac{\operatorname{se}_{\mathrm{full}}}
 {\operatorname{se}_{\mathrm{GN}}}\longrightarrow\frac qA.
 \label{eq:se-ratio}
\end{equation}
The same consistency result gives $\widehat q/\widehat A\to_p q/A$. On regular samples with $\widehat A>0$ and a common positive score-variance estimate, $\widehat q/\widehat A$ is exactly the ratio of the two fitted standard errors, irrespective of the HAC bandwidth. The quantity $\sqrt{\Omega/T}/A$ gives a first-order standard-deviation approximation. Convergence of finite-sample second moments additionally requires uniform integrability.

Correct specification of the fitted $K$-lag vector is sufficient for $A=q$,
but is not necessary in general. The exact condition is
$\sum_kR_km_k''(p^\star)=0$; nonzero residuals alone do not prove inequality.
For $K=1$, $A=q=1$ for every admissible process. Appendix~\ref{app:equality} gives a valid $K=3$ example with nonzero residuals and $A=q$.
Estimation of the shared variance denominator contributes the term $-\rho_kY_t^2$. The first-order condition converts the weighted sum of these terms to \eqref{eq:influence-score}.
The disagreement with the general AD variance expression in
\cite[Eq.~17]{Riegner2026} concerns the Gauss--Newton denominator when
residual curvature is nonzero.

For $p^\star\neq0$, $g(p)=-1/\log|p|$ is smooth near the target and
\begin{equation}
 g'(p)=\frac1{p\{\log|p|\}^2},\qquad
 \sqrt T\{g(\widehat p)-g(p^\star)\}
 \xrightarrow{d}N\!\left(0,\frac{g'(p^\star)^2\Omega}{A^2}\right).
 \label{eq:tau-delta}
\end{equation}
Multiplication by the sampling interval converts the target and standard errors to physical time units and preserves Wald coverage exactly. Finite-sample coverage may differ between the $p$ and $g(p)$ scales because $g$ is nonlinear.

\subsection{The implementation's lagged-square score}
The implementation-compatible population score is
\begin{equation}
 v_t=\frac1{\gamma_0}\sum_{k=1}^K b_k
 \{Y_tY_{t-k}-(p^\star)^kY_{t-k}^2\}.
 \label{eq:lagged-score}
\end{equation}
Its difference from \eqref{eq:influence-score} is
\begin{equation}
 v_t-u_t=\frac1{\gamma_0}\sum_{k=1}^K b_k(p^\star)^k
 (Y_t^2-Y_{t-k}^2).
 \label{eq:score-difference}
\end{equation}
For fixed $K$, the time sum consists of finitely many boundary terms,
so $T^{-1/2}\sum_t(v_t-u_t)=o_p(1)$. Under absolute covariance
summability the scores have the same long-run variance $\Omega$.
They can nevertheless have different lag-specific covariances and different
finite-bandwidth Bartlett functionals. The bandwidth calculations use $v_t$.

\section{Error direction and regular counterexamples}
\label{sec:direction}
\begin{proposition}[A conditional single-crossing rule]
\label{prop:single-crossing}
Suppose $p^\star>0$, $A>0$, and the first-order condition holds.
If, for some $j\in\{1,\ldots,K-1\}$, $R_k\leq0$ for $k\leq j$ and
$R_k\geq0$ for $k>j$, then $q\geq A$. The inequality is strict for a
nonzero residual vector. Reversing the signs gives $q\leq A$, also
strictly for a nonzero residual vector.
\end{proposition}
\begin{proof}
Since $b_k>0$ and $m_k''(p^\star)=(k-1)b_k/p^\star$,
the first-order condition gives, for any real $c_0$,
\[
 q-A=\frac1{p^\star}\sum_k(k-1-c_0)b_kR_k.
\]
Take $c_0=j-1/2$. Every term is nonnegative under the first sign
pattern, with at least one positive term for a nonzero residual vector.
Reverse the signs for the second assertion.
\end{proof}
Application to an oscillatory process requires verification of the stated residual sign condition.

Consider the variance-one Gaussian observation-noise family
\begin{equation}
 X_t=\sqrt c\,Z_t+\sqrt{1-c}\,\varepsilon_t,\qquad
 Z_t=aZ_{t-1}+\sqrt{1-a^2}\,\eta_t,
 \label{eq:noise-family}
\end{equation}
where $0<a<1$, $0<c<1$, the innovation sequences are independent standard
normal white noises, and $Z_t$ is stationary. Then $\rho_k=ca^k$ for
$k\geq1$. For this family take $\mathcal P=[-r,r]$ with $a<r<1$.
At a regular interior minimizer with $K\geq2$, $0<p^\star<a$,
and $R_k/(p^\star)^k=c(a/p^\star)^k-1$ is strictly increasing in $k$.
The first-order condition forces both signs, giving $A<q$ by
Proposition~\ref{prop:single-crossing}. Appendix~\ref{app:direction-family}
gives the details and the sign-symmetry argument for negative $a$.

With $a=3/4$, $c=4/7$, and $K=2$, the exact values are
\begin{equation}
 p^\star=\frac12,\quad q=2,\quad A=\frac{13}{7},\quad
 \Omega=\frac{52963}{10976},\quad \frac qA=\frac{14}{13}.
 \label{eq:exact-noise}
\end{equation}
A counterexample in the opposite direction is the Gaussian MA(1) process
\begin{equation}
 X_t=\frac{\varepsilon_t+\theta\varepsilon_{t-1}}{\sqrt{1+\theta^2}},
 \qquad \theta=\frac{27-\sqrt{245}}{22}.
 \label{eq:ma-example}
\end{equation}
It has $\rho_1=11/27$ and $\rho_k=0$ for $k\geq2$. For $K=2$,
\begin{equation}
 p^\star=\frac13,\quad q=\frac{13}{9},\quad A=\frac53,\quad
 \Omega=\frac{1121518}{531441},\quad \frac qA=\frac{13}{15}.
 \label{eq:exact-ma}
\end{equation}
Both examples define stationary Gaussian processes with unique regular interior projection minima and positive long-run variances. Exact calculations are given in Appendix~\ref{app:exact-examples}.

With an otherwise consistent long-run variance estimate, a nominal $95\%$
GN Wald interval has limiting coverage
\begin{equation}
 2\Phi\!\left(z_{0.975}\frac Aq\right)-1.
 \label{eq:gn-limit}
\end{equation}
The analytically calculated coverage limits are $93.1236\%$ in \eqref{eq:exact-noise} and $97.6271\%$ in \eqref{eq:exact-ma}. Thus the first-order discrepancy can have either direction.

\section{Fixed-bandwidth HAC and Wald coverage}
\label{sec:fixed-m}
Curvature and long-run variance estimation are separate parts of the sandwich.
Let $n=T-K$, and let $\widehat v_t$, $t=K+1,\ldots,T$, denote
\eqref{eq:lagged-score} evaluated at the sample estimates of $p^\star$,
$\mu$, and $\gamma_0$. For a fixed integer $0\leq M<n$, the noncentered
Bartlett estimator of the score-mean variance is
\begin{equation}
 \widehat V_M=\frac1{n^2}\left\{
 \sum_{t=K+1}^T\widehat v_t^2+
 2\sum_{h=1}^M\left(1-\frac{h}{M+1}\right)
 \sum_{t=K+1+h}^T\widehat v_t\widehat v_{t-h}
 \right\}.
 \label{eq:nw-v}
\end{equation}
This retains the normalization and lagged-square score of the implementation.
The full and GN standard errors on regular samples are respectively
$\sqrt{\widehat V_M}/\widehat A$ and $\sqrt{\widehat V_M}/\widehat q$.
Let
\begin{equation}
 c_v(h)=\operatorname{Cov}(v_t,v_{t-h}),\qquad
 \Omega_M=c_v(0)+2\sum_{h=1}^M
 \left(1-\frac{h}{M+1}\right)c_v(h),\qquad r_M=\frac{\Omega_M}{\Omega}.
 \label{eq:bartlett-population}
\end{equation}
At the population target $\mathbb E v_t=0$; the sample estimator nevertheless
retains the original noncentered products.

\begin{proposition}[Coverage with fixed $K$ and $M$]
\label{prop:fixed-m}
Under Assumption~\ref{ass:fixed-k}, with fixed $K,M$ and $\Omega_M>0$,
$n\widehat V_M\to_p\Omega_M$. The limiting coverages of the full and
GN nominal $1-\alpha$ normal Wald intervals are
\begin{align}
 C_{\mathrm{full}}(M)
 &=2\Phi\!\left(z_{1-\alpha/2}\sqrt{r_M}\right)-1,
 \label{eq:full-fixed-m}\\
 C_{\mathrm{GN}}(M)
 &=2\Phi\!\left(z_{1-\alpha/2}\frac Aq\sqrt{r_M}\right)-1.
 \label{eq:gn-fixed-m}
\end{align}
For $p^\star\neq0$, the delta-method Wald intervals for $g(p^\star)$
have the same respective coverage limits.
\end{proposition}
\begin{proof}
For fixed $K,h$, $\widehat v_t\widehat v_{t-h}$ is a finite linear
combination of products of up to four observations, with coefficients
continuous in the consistent estimates of $p^\star$, $\mu$, and
$\gamma_0>0$. Ergodicity and the fourth-moment condition give convergence
of the empirical averages to $\mathbb E(v_tv_{t-h})=c_v(h)$.
Omitting finitely many boundary terms has no effect. Summing over the
fixed set $h\leq M$ gives $n\widehat V_M\to_p\Omega_M$.
Because $T/n\to1$, Slutsky's theorem and Theorem~\ref{thm:asymptotic}
give both coverage limits. The smooth, nonzero derivative in
\eqref{eq:tau-delta} cancels in the limiting studentized statistic.
\end{proof}

When $r_M<1$, the limiting coverage of the full-Hessian interval remains below nominal. Standard HAC consistency regimes combine increasing bandwidth and a vanishing relative window with additional dependence, moment, and plug-in regularity conditions \cite{NeweyWest1987}.

The results hold for fixed $K$. Extending them to $K=\lfloor T/100\rfloor$ requires control of the changing projection target, score, and boundary terms. For fixed $K$, $\Omega/(T-K)$ and $\Omega/T$ are first-order equivalent variance approximations.

\section{Synthetic experiments and analysis conventions}\label{sec:methods}
\subsection{Simulation design}
The simulations comprised a pilot and a pilot-informed follow-up study, analyzed separately. Pilot~A contains 16,000 unique Gaussian series: two models at $K=2$, $T\in\{500,2000,8000\}$ with 2,000 repetitions per cell, followed by two $K=36,T=3600$ cells with 2,000 repetitions each. For the $K=36$ cells, the same series were evaluated at $M=5$ and $M=8$.

The follow-up, denoted A2, contains 24,000 new series in eight cells. At $K=36,T=3600$, four models each have $R=4000$ repetitions: exact AR1 with coefficient $0.5$; exact AR1 with coefficient $0.75$; an AR1-plus-independent-noise mixture with $(a,c)=(0.75,4/7)$; and a more persistent mixture with $(a,c)=(0.9,4/7)$. The first and third models additionally have $T=1000$ and $T=10000$ cells with $R=2000$ each. The mixture is
\begin{equation}
 X_t=\sqrt c\,Z_t+\sqrt{1-c}\,\varepsilon_t,
 \qquad \Var(Z_t)=\Var(\varepsilon_t)=1,
 \qquad \operatorname{Corr}(Z_t,Z_{t-k})=a^k,
 \label{eq:simulation-family}
\end{equation}
where $Z_t$ is a stationary Gaussian AR1 process and $\varepsilon_t$ is independent Gaussian white noise. Initial states are drawn from the stationary distribution. The exact-$0.75$ control matches the latent persistence $a$ of the main mixture, not its projection parameter $p^*$. The models isolate temporal dependence and observation noise under tractable Gaussian assumptions.

A2 used a pilot-informed design with an analysis plan recorded locally before execution; the study was not publicly preregistered. All specified cells, bandwidths, and arms are reported. The designated primary A2 comparison is full versus Gauss--Newton \emph{sample} bread with the same HAC estimate, for $\phi=p$ coverage in the $(0.75,4/7),T=3600,M=5$ cell. The transformed timescale $\tau$ is secondary. The subsequent variability, nesting, directional-miss, and source-process analyses were post hoc and used saved outputs or deterministic calculations.

\subsection{Common fit and common HAC estimate}
Each series is centered over its full length and standardized with the population-denominator sample variance (\texttt{ddof=0}). The sample ACF uses the full lag-zero sum in the denominator and the available lag-$k$ product sum in the numerator. We minimize the same unweighted ACF objective for all uncertainty calculations. The implementation enumerates numerically real stationary roots of its derivative polynomial and compares their objective values with those at the endpoints $\pm(1-10^{-10})$. This optimizer differs from the constrained \texttt{curve\_fit} routine in the public implementation identified in Appendix~\ref{app:reference-code}. At $K=36$ the derivative polynomial has degree 71.

Both bread choices use the same fitted parameter and the same lagged-square score convention as the public implementation. With $n=T-K$, fitted scores $\hat v_t$ on the common window, and bandwidth $M$, the score-mean variance estimate is
\begin{equation}
 \widehat V_M=\frac{1}{n^2}\left\{
 \sum_{t=1}^{n}\hat v_t^2+
 2\sum_{h=1}^{M}\left(1-\frac{h}{M+1}\right)
 \sum_{t=h+1}^{n}\hat v_t\hat v_{t-h}\right\}.
 \label{eq:sample-hac}
\end{equation}
Scores are not additionally demeaned. A2 evaluates $M\in\{5,8,16,32,64\}$ on the same series. The primary bandwidth $M=5$ follows the public AD analysis notebook; $M=8,16,32,64$ are prespecified sensitivity settings.

For every cell, we flag nonfinite values, curvature at or below $10^{-10}$, a stationary-equation residual of at least $10^{-7}$, and fitted values within $10^{-6}$ of zero or either unit boundary. All-repetition coverage treats invalid replicates as noncoverage; valid-only coverage is also retained. No fit or HAC failures occurred in the evaluated cells.

\subsection{Factorial diagnostics and endpoints}
The eight A2 arms cross (i) bread form, $q$ or $A$; (ii) bread source, sample or population; and (iii) score-mean variance source, $\widehat V_M$ or a population oracle. The primary A2 oracle convention is $\Omega/n$, to match the score-window denominator, while $\Omega/T$ is retained as a sensitivity corresponding directly to the fixed-$K$ expansion. Both are first-order diagnostics, not the exact finite-sample variance of the fitted estimator. Oracle quantities require knowledge of the generating model and are not feasible competitors for practical use.

Parameter intervals are unclipped normal Wald intervals. Timescale intervals use $\hat\tau=-1/\log|\hat p|$ and the absolute delta-method derivative $1/(|\hat p|\log^2|\hat p|)$. Coverage was evaluated separately for the parameter and timescale intervals. A marginal coverage estimate has a Wilson Monte Carlo (MC) interval; paired differences use the sample standard deviation of $D_r=I_{r,\mathrm{full}}-I_{r,\mathrm{GN}}$ divided by $\sqrt R$. Factorial contrasts and interactions are evaluated on these shared indicators. Their changes are conditional substitution effects, not additive percentages of the error attributable to individual components. MC intervals are condition-specific and are not simultaneous inferential guarantees over all diagnostic cells.

\subsection{Exploratory analyses}
Post hoc analyses of the saved replicate outputs compared the empirical standard deviation of $\hat p$ with $\sqrt{\Omega/T}/A$ and $\sqrt{\Omega/T}/q$, alongside the existing $n$ convention. These comparisons use the population LRV and isolate the scale implication of the bread from HAC estimation. We also computed the fraction of nested intervals, the two directional miss probabilities, and correlations between estimates and standard errors. These correlations describe associations rather than causal effects; correlations involving a constant oracle standard error are undefined.

The observation-noise and stationary AR2 specifications in \citet{Riegner2026} were additionally evaluated by deterministic AD projection at $K=36$ and $48$. The original experiments evaluated TD estimation. Calculations use the printed parameter values and report finite-$K$ sensitivity.

\section{Results}\label{sec:results}
\subsection{Sampling variability and population first-order scales}
Write $s_{A,T}=\sqrt{\Omega/T}/A$ and $s_{q,T}=\sqrt{\Omega/T}/q$. These are known-model reference scales, not fitted HAC standard errors. At $T=3600$, empirical SD divided by $s_{A,T}$ ranges from $0.990$ to $1.009$ across the four models (\cref{tab:sd,fig:sd}). In the exactly specified controls the two population scales coincide. In the noise mixtures, empirical SD is about $1.327$ and $1.451$ times the Gauss--Newton scale, respectively. These gaps are much larger than the approximately one-percent MC uncertainty of an SD estimate with $R=4000$.

\begin{table}[tbp]\centering\small
\caption{Empirical variability and population first-order scales, $K=36$. The $T$ convention follows the asymptotic expansion. All eight cells are shown. The comparison is exploratory.}\label{tab:sd}
\begin{tabular}{lrrrrrr}
\toprule
Process and length & $R$ & Empirical SD & $s_{A,T}$ & $s_{q,T}$ & $s/s_{A,T}$ & $s/s_{q,T}$ \\
\midrule
AR1 $(.5)$, $T=3600$ & 4000 & 0.01976 & 0.01963 & 0.01963 & 1.007 & 1.007 \\
AR1 $(.75)$, $T=3600$ & 4000 & 0.01809 & 0.01792 & 0.01792 & 1.009 & 1.009 \\
AR1 $(.75)$+noise, $T=3600$ & 4000 & 0.02748 & 0.02775 & 0.02071 & 0.990 & 1.327 \\
AR1 $(.9)$+noise, $T=3600$ & 4000 & 0.02415 & 0.02414 & 0.01664 & 1.001 & 1.451 \\
AR1 $(.5)$, $T=1000$ & 2000 & 0.03624 & 0.03725 & 0.03725 & 0.973 & 0.973 \\
AR1 $(.5)$, $T=10000$ & 2000 & 0.01168 & 0.01178 & 0.01178 & 0.991 & 0.991 \\
AR1 $(.75)$+noise, $T=1000$ & 2000 & 0.05256 & 0.05265 & 0.03929 & 0.998 & 1.338 \\
AR1 $(.75)$+noise, $T=10000$ & 2000 & 0.01687 & 0.01665 & 0.01242 & 1.013 & 1.358 \\
\bottomrule
\end{tabular}

\end{table}
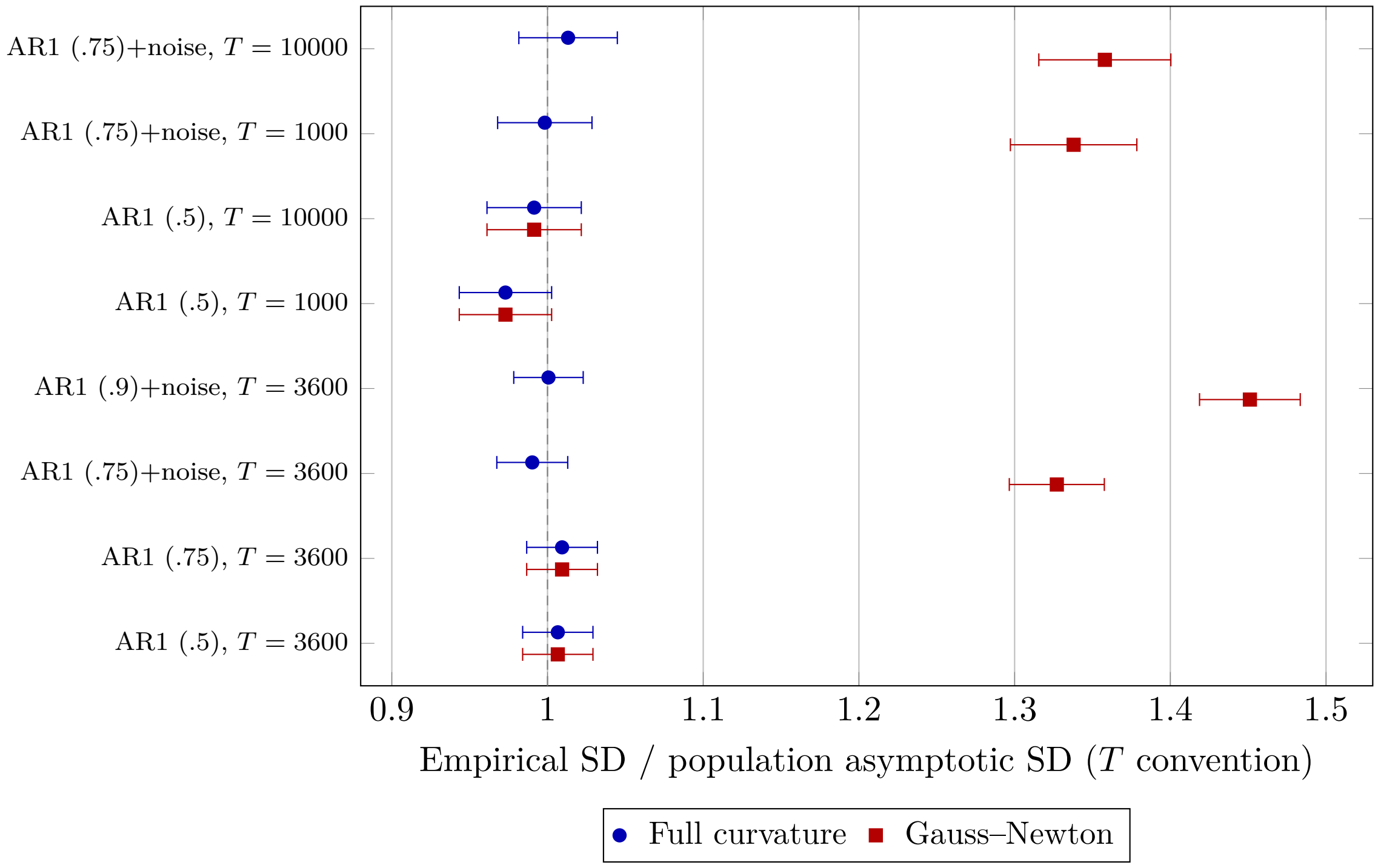
\begin{figure}[tbp]\centering
\begin{tikzpicture}
\begin{axis}[width=0.82\linewidth,height=9cm,xmin=0.88,xmax=1.53,ymin=-0.5,ymax=7.5,ytick={0,1,2,3,4,5,6,7},yticklabels={{AR1 $(.5)$, $T=3600$},{AR1 $(.75)$, $T=3600$},{AR1 $(.75)$+noise, $T=3600$},{AR1 $(.9)$+noise, $T=3600$},{AR1 $(.5)$, $T=1000$},{AR1 $(.5)$, $T=10000$},{AR1 $(.75)$+noise, $T=1000$},{AR1 $(.75)$+noise, $T=10000$}},xlabel={Empirical SD / population asymptotic SD ($T$ convention)},xmajorgrids=true,yticklabel style={font=\scriptsize},legend style={at={(0.5,-0.18)},anchor=north,legend columns=2,column sep=5pt,font=\small}]
\addplot[gray,densely dashed,forget plot] coordinates {(1,-.5) (1,7.5)};
\addplot+[only marks,color=blue!70!black,mark=*,mark options={fill=blue!70!black},error bars/.cd,x dir=both,x explicit] coordinates {(1.00661357,0.13) +- (0.02257786,0) (1.00935709,1.13) +- (0.02272094,0) (0.99020984,2.13) +- (0.02277211,0) (1.00062309,3.13) +- (0.02230193,0) (0.97298500,4.13) +- (0.02966095,0) (0.99141580,5.13) +- (0.03028344,0) (0.99826209,6.13) +- (0.03029626,0) (1.01322182,7.13) +- (0.03163265,0)};
\addlegendentry{Full curvature}
\addplot+[only marks,color=red!70!black,mark=square*,mark options={fill=red!70!black},error bars/.cd,x dir=both,x explicit] coordinates {(1.00661357,-0.13) +- (0.02257786,0) (1.00935709,0.87) +- (0.02272094,0) (1.32710718,1.87) +- (0.03051983,0) (1.45111325,2.87) +- (0.03234247,0) (0.97298500,3.87) +- (0.02966095,0) (0.99141580,4.87) +- (0.03028344,0) (1.33789903,5.87) +- (0.04060390,0) (1.35794848,6.87) +- (0.04239498,0)};
\addlegendentry{Gauss--Newton}
\end{axis}
\end{tikzpicture}
\caption{Empirical SD divided by the full-curvature or Gauss--Newton population scale. The vertical line marks equality. Bars are $\pm1.96$ estimated MC standard errors, obtained by a delta calculation using the fourth empirical central moment of the replicate estimates. The bars quantify Monte Carlo uncertainty in the empirical-SD ratios.}\label{fig:sd}
\end{figure}
Across all eight cells the empirical/full ratios range from $0.973$ to $1.013$. The original A2 $n=T-K$ convention is also tabulated in the appendix. Agreement in an SD summary does not establish normal tails or balanced two-sided calibration.

\subsection{The designated coverage comparison and nesting}
In the designated mixture cell with $T=3600,M=5$, parameter coverage increased from $77.825\%$ to $90.500\%$ after replacing sample $q$ with sample $A$. The paired difference is $12.675$ percentage points (MC $95\%$ interval $[11.644,13.706]$). However, all 4,000 fitted ratios $\hat q/\hat A$ exceeded one: every full-curvature interval contained its Gauss--Newton counterpart. The gain is therefore structurally constrained to be nonnegative. This nesting limits the use of the coverage gain as evidence for the variance formula; the population-scale comparison evaluates that formula directly.

Full-curvature coverage remains below $95\%$ (Wilson MC interval $[89.552,91.370]\%$). For the timescale endpoint the rates are $78.200\%$ and $89.675\%$. The parameter and timescale intervals therefore have different finite-sample coverage.

\subsection{Fixed-bandwidth limits and remaining discrepancies}
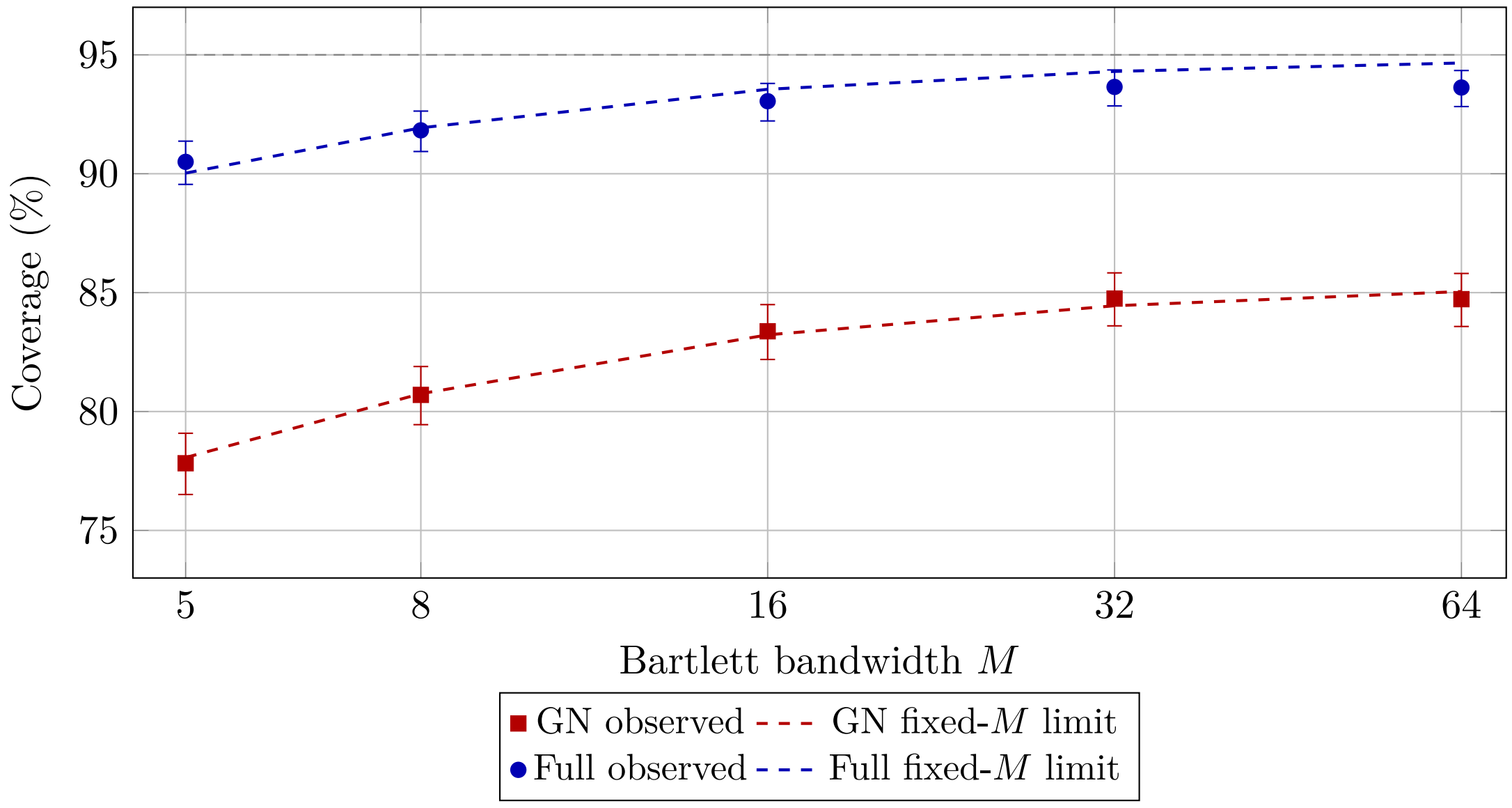
\begin{figure}[tbp]\centering
\begin{tikzpicture}
\begin{axis}[width=0.95\linewidth,height=7cm,xmode=log,log basis x=2,xtick={5,8,16,32,64},xticklabels={5,8,16,32,64},xmin=4.5,xmax=70,ymin=73,ymax=97,xlabel={Bartlett bandwidth $M$},ylabel={Coverage (\%)},grid=major,legend style={at={(0.5,-0.2)},anchor=north,legend columns=2,font=\small}]
\addplot[gray,densely dashed,forget plot] coordinates {(5,95) (64,95)};
\addplot+[only marks,color=red!70!black,mark=square*,mark options={fill=red!70!black},error bars/.cd,y dir=both,y explicit] coordinates {(5,77.82500) += (0,1.26035) -= (0,1.31374) (8,80.70000) += (0,1.19333) -= (0,1.25224) (16,83.37500) += (0,1.12163) -= (0,1.18568) (32,84.75000) += (0,1.08072) -= (0,1.14740) (64,84.72500) += (0,1.08149) -= (0,1.14812)};
\addlegendentry{GN observed}
\addplot[color=red!70!black,thick,dashed] coordinates {(5,78.064391) (8,80.751130) (16,83.225216) (32,84.447571) (64,85.046571)};
\addlegendentry{GN fixed-$M$ limit}
\addplot+[only marks,color=blue!70!black,mark=*,mark options={fill=blue!70!black},error bars/.cd,y dir=both,y explicit] coordinates {(5,90.50000) += (0,0.87020) -= (0,0.94792) (8,91.82500) += (0,0.80948) -= (0,0.88974) (16,93.05000) += (0,0.74748) -= (0,0.83009) (32,93.65000) += (0,0.71463) -= (0,0.79839) (64,93.62500) += (0,0.71604) -= (0,0.79975)};
\addlegendentry{Full observed}
\addplot[color=blue!70!black,thick,dashed] coordinates {(5,90.026192) (8,91.930285) (16,93.551526) (32,94.302571) (64,94.657973)};
\addlegendentry{Full fixed-$M$ limit}
\end{axis}
\end{tikzpicture}
\caption{Parameter coverage in the canonical $T=3600$ mixture and the fixed-$K$, fixed-$M$ limits. Dashed colored lines are deterministic limits for the implementation-compatible lagged-square score; points and Wilson MC intervals use the same 4,000 series across all bandwidths. The gray line denotes $95\%$. Finite-sample departures from a limit are not attributed to a single cause by this comparison.}\label{fig:calibration}
\end{figure}

The canonical mixture has $\Omega_5/\Omega=0.705390$. Its full-curvature fixed-$M$ coverage limit is $90.026\%$, whereas the additional Gauss--Newton denominator discrepancy gives $78.064\%$ (\cref{fig:calibration}). Some undercoverage therefore persists when $T$ increases with $M$ fixed; it is not solely a finite-sample error.
For $M=5,8,16,32,64$, the full-curvature limits are $90.026$, $91.930$, $93.552$, $94.303$, and $94.658\%$; the observed rates are $90.500$, $91.825$, $93.050$, $93.650$, and $93.625\%$. At larger bandwidths the observations remain below their limits. HAC variability, plug-in dependence, estimator bias, and nonnormality may contribute; these data do not identify a unique decomposition.

\subsection{Conditional substitutions and directional misses}
At the same sample $A$, replacing HAC with $\Omega/n$ changes primary-cell parameter coverage from $90.500\%$ to $94.900\%$. In contrast, replacing sample $A$ with population $A$ while retaining HAC changes it to $89.100\%$. Using both population $A$ and oracle variance gives $95.275\%$. These are conditional substitution effects, not additive error contributions.
\begin{table}[tbp]\centering\small
\caption{Exploratory directional-miss analysis in the canonical $T=3600,M=5$ cell. Below lower means $p^*$ is below the lower endpoint; above upper means it is above the upper endpoint. Rates are percentages. Oracle uses $\Omega/n$. A dash indicates an undefined correlation for a constant SE.}\label{tab:tails}
\begin{tabular}{lrrrr}
\toprule
Bread / score variance & Coverage & Below lower & Above upper & $\operatorname{cor}(\hat p,\mathrm{SE})$ \\
\midrule
Sample $A$ / HAC & 90.500 & 4.050 & 5.450 & -0.18 \\
Population $A$ / HAC & 89.100 & 0.400 & 10.500 & 0.94 \\
Sample $A$ / oracle & 94.900 & 4.700 & 0.400 & -0.95 \\
Population $A$ / oracle & 95.275 & 1.875 & 2.850 & -- \\
\bottomrule
\end{tabular}

\end{table}
The $94.900\%$ sample-$A$/oracle rate masks misses of $4.700\%$ below the lower endpoint and $0.400\%$ above the upper endpoint (\cref{tab:tails}). A total near $95\%$ need not imply balanced two-sided calibration. The estimate--SE correlations describe associated changes in interval width and location.

\subsection{Adverse controls, longer series, and the timescale endpoint}
Full sample curvature is not uniformly beneficial. For exact AR1$(0.5)$ at $T=1000,K=36,M=5$, parameter coverage decreases from $93.750\%$ to $92.300\%$, a paired difference of $-1.450$ percentage points (MC interval $[-2.174,-0.726]$). At $M=64$ it decreases from $92.650\%$ to $91.450\%$. Population $A=q$ here, but sample residual curvature is variable. These results are specific to $K=36$; the source package's default $K=\lfloor T/100\rfloor$ would give $K=10$ and was not evaluated in these cells.

Exact AR1$(0.75)$ at $T=3600,M=5$ also undercovers ($88.225\%$ GN and $87.800\%$ full). Its population $\Omega_5/\Omega=0.658356$ shows a short-bandwidth limitation without a population bread discrepancy. This does not attribute all finite-sample error to HAC. In the persistent mixture, full parameter coverage is $77.775\%$ at $M=5$ and $92.250\%$ at $M=64$.

For the canonical mixture at $T=10000$, full $M=64$ coverage is $92.950\%$. Even population-$A$/oracle-$n$ coverage is $93.900\%$ (Wilson interval $[92.765,94.867]\%$), versus $93.850\%$ using $\Omega/T$. Across the eight cells, population-full/oracle-$n$ rates span $93.900$--$96.500\%$. These condition-specific MC intervals are unadjusted for comparisons across diagnostic cells. In the persistent mixture, population-full/oracle-$n$ coverage is $94.675\%$ for $p$ but $91.850\%$ for $\tau$, showing a further delta-Wald limitation.

\subsection{AD projections of published process specifications}
The noise process in Eq.~(22) of \citet{Riegner2026} has latent AR1 coefficient $0.8$ and innovation variance one, giving latent variance $1/(1-0.8^2)$. Added noise variance $\sigma^2$ thus implies $c=[1+(1-0.8^2)\sigma^2]^{-1}$.

For the stationary AR2 specifications with TD target $0.8$, we use $\phi_2=-r^2$ and $\phi_1=0.8(1+r^2)$, treating the printed radii as exact. The $r=0.5$ case is a critically damped repeated-root boundary; the other radii are oscillatory. The source article evaluated these process designs for TD estimation. Table~\ref{tab:source-designs} reports the corresponding deterministic AD projections.
\begin{table}[tbp]\centering\small
\caption{Deterministic AD projection of published process specifications. Noise parameter: $\sigma^2$; AR2 parameter: $r=\sqrt{-\phi_2}$. Printed values are treated as exact inputs. GN limits assume a consistent common LRV and use $K=48$; they are not finite-sample observations. $K=36$ and 48 values are close but not necessarily identical.}\label{tab:source-designs}
\begin{tabular}{lrrrrr}
\toprule
Process & Parameter & $p^*_{48}$ & $q/A$ ($K=36$) & $q/A$ ($K=48$) & GN limit (\%) \\
\midrule
Noise & 0.000 & 0.80000 & 1.00000 & 1.00000 & 95.000 \\
Noise & 0.500 & 0.76776 & 1.08782 & 1.08783 & 92.841 \\
Noise & 1.000 & 0.73414 & 1.17960 & 1.17960 & 90.340 \\
Noise & 1.500 & 0.69909 & 1.27394 & 1.27394 & 87.607 \\
Noise & 2.000 & 0.66270 & 1.36821 & 1.36821 & 84.800 \\
AR2 & 0.500 & 0.69738 & 0.88311 & 0.88311 & 97.354 \\
AR2 & 0.642 & 0.61878 & 0.74759 & 0.74759 & 99.125 \\
AR2 & 0.758 & 0.54064 & 0.59700 & 0.59700 & 99.897 \\
AR2 & 0.859 & 0.47109 & 0.46946 & 0.46946 & 99.997 \\
AR2 & 0.949 & 0.41360 & 0.37632 & 0.37632 & $>99.999$ \\
\bottomrule
\end{tabular}

\end{table}
The positive-noise-variance cases have $q/A>1$, while the zero-noise control has $q/A=1$ and the evaluated AR2 grid has $q/A<1$. Several AR2 residual sequences change sign repeatedly, so the single-crossing lemma does not establish their direction. Their signs follow from direct curvature calculations. Similarly, numerical stability of $q/A$ as $K$ increases is not a growing-$K$ proof.

\section{Discussion}\label{sec:discussion}
\subsection{Implications for autocorrelation-based inference}
The full sensitivity of the fixed-$K$ ACF estimating equation includes a residual-curvature term. This standard component of misspecified extremum estimation can be consequential for the AD variance formula of \citet{Riegner2026}. The stationary Gaussian examples demonstrate both underestimation and overestimation of the first-order standard error when the term is omitted. Exact specification is sufficient for $A=q$; more generally, equality holds whenever the residual-curvature inner product vanishes. At $K=1$, equality holds for every admissible process. Under the fixed-$K$ assumptions, the fitted ratio $\widehat q/\widehat A$ consistently estimates the curvature contribution to the standard-error ratio. It does not assess long-run variance estimation or overall interval calibration, may be unstable when the fitted curvature is small, and cannot establish correct specification when it is near one.

The analytical examples and empirical-SD comparisons assess the first-order scale independently of HAC estimation. Wald coverage also depends on estimation of the standard error and the distribution of the studentized statistic. In all 4,000 observed replicates of the primary mixture condition, full-curvature intervals contained their Gauss--Newton counterparts: they shared the same center and HAC estimate and had larger fitted standard errors. This nesting constrained the observed coverage gain to be nonnegative. In the exact-model control, sample residual curvature can instead reduce coverage. Thus a correct population curvature does not ensure superior finite-sample performance for every plug-in interval.

Fixed-bandwidth HAC introduces a separate discrepancy because it targets $\Omega_M$. Consistent estimation of $\Omega$ requires a suitable increasing-bandwidth regime and additional regularity conditions. Evaluation of a practical interval procedure must therefore address bandwidth selection, sample-curvature variability, and the transformation to the timescale endpoint jointly. Comparisons with adaptive-bandwidth and bootstrap procedures would help assess finite-sample calibration beyond the fixed procedures considered here.

\subsection{Limitations}
The simulations use four Gaussian processes with tractable LRV and Bartlett calculations. Performance under non-Gaussianity, censoring, motion, hemodynamic convolution, nonstationarity, and spatial multiplicity remains unassessed. The study contains no analysis of empirical fMRI maps or downstream imaging conclusions. Its residual-curvature result concerns the nonlinear AD projection; the TD linear projection has a different sensitivity.

All asymptotic results assume fixed $K$. The source package's default $K=\lfloor T/100\rfloor$ requires additional control of boundary terms, normalization, and the changing projection target. Stability of the computed population ratios does not establish that extension. The deterministic AR2 results apply to the specified parameter grid and cannot be generalized to all damped oscillatory processes.

The follow-up design was informed by a pilot, and the variability, directional-miss, and source-process analyses were conducted post hoc. These diagnostics are exploratory. Raw simulated series and lag-specific sample ACF vectors were not retained; the independent computational audit therefore validates aggregation of saved replicate outputs rather than regenerating the underlying series.

\subsection{Conclusion}
For fixed-lag exponential-ACF projection, residual curvature is part of the estimating-equation sensitivity. Omitting it can shift a Gauss--Newton standard error above or below the correct first-order scale. Full-curvature substitution resolves this denominator discrepancy, while bandwidth estimation, finite-sample studentization, and the nonlinear timescale transformation remain determinants of Wald calibration.

\section*{Data and code availability}
The companion archive (\url{https://doi.org/10.5281/zenodo.24283730}) contains the pilot and follow-up analysis plans, code, configurations, saved replicate outputs, and SHA-256 manifests. The follow-up simulations used Python~3.12.14, NumPy~2.3.5, PCG64 with disjoint cell-specific streams, and single-threaded BLAS settings. The seed was 20261116015702 and the batch size was 128; the configuration specifies the generation order. All data are synthetic.

Independent reaggregation of the saved outputs agreed with all 18,911 numerical summaries to within $8.89\times10^{-16}$. Raw simulated series and lag-specific sample ACF vectors are unavailable.

\section*{Declaration of generative AI use}
GPT-6 Astra (OpenAI) assisted with study design, code development and implementation, and review. Claude Fable 5.2 (Anthropic) assisted with independent review, comparison with source texts and public code, mathematical derivations, and numerical cross-checking. The author takes responsibility for the scientific claims and final content of the manuscript.

\section*{Acknowledgments and funding}
The author acknowledges OpenAI for providing complimentary access to ChatGPT through the ChatGPT for Academic Researchers program. No other funding was received for this work.

\section*{Competing interests}
The author received the complimentary access described above and declares no other competing interests.

\begingroup
\raggedright
\bibliographystyle{vancouver-timescale}
\bibliography{references}
\endgroup
\appendix
\section{Moment expansion and Gaussian calculations}
\label{app:derivations}

\subsection{Centering, boundary terms, and the ACF ratio}
\label{app:linearization}
Define
\[
 \widetilde\gamma_k=T^{-1}\sum_{t=1}^T Y_tY_{t-k},\qquad
 \widehat\gamma_k=T^{-1}\sum_{t=k+1}^T
 (Y_t-\bar Y_T)(Y_{t-k}-\bar Y_T),\quad 1\leq k\leq K,
\]
and $\widehat\gamma_0=T^{-1}\sum_{t=1}^T(Y_t-\bar Y_T)^2$.
The stationary extension to unobserved negative times introduces finitely many boundary products, whose sum is $O_p(1)$.
For fixed $k$, each truncated sample mean differs from $\bar Y_T$
by $O_p(T^{-1})$. The assumption $\bar Y_T=O_p(T^{-1/2})$ therefore yields
\[
 \widehat\gamma_k-\widetilde\gamma_k=O_p(T^{-1}),
 \qquad k=0,\ldots,K.
\]
For $k=0$ the difference is exactly $-\bar Y_T^2$.
The delta method for $\widehat\rho_k=\widehat\gamma_k/\widehat\gamma_0$
then gives
\[
 \sqrt T(\widehat\rho_k-\rho_k)=
 \gamma_0^{-1}\sqrt T\{(\widetilde\gamma_k-\gamma_k)
 -\rho_k(\widetilde\gamma_0-\gamma_0)\}+o_p(1).
\]
In matrix notation the ACF covariance limit is
\[
 \Sigma_\rho=D\Sigma_VD^\top,\qquad
 D=\gamma_0^{-1}[-\rho\ \ I_K],\qquad
 \Omega=b^\top\Sigma_\rho b,
\]
where $\rho=(\rho_1,\ldots,\rho_K)^\top$ and $b=(b_1,\ldots,b_K)^\top$.
This ratio delta-method representation accounts for both autocovariance and variance estimation and is consistent with the classical treatment of sample autocorrelations under weak dependence \citep{RomanoThombs1996}. The covariance $\Sigma_V$ retains the fourth-order dependence of the process; a covariance-only Gaussian formula requires the additional Gaussian structure used below.

Consistency puts the sample minimizer in the interior with probability
tending to one. The ordinary first-order condition then applies. A numerical
approximation can be treated by the same argument if its score residual is
$o_p(T^{-1/2})$ and it selects the consistent minimizer; numerical diagnostics
in a finite run do not by themselves prove this asymptotic property.

\subsection{Long-run variance of a Gaussian quadratic score}
\label{app:gaussian-lrv}
For a mean-zero stationary Gaussian process, set
\[
 b_0=-\sum_{k=1}^K b_k(p^\star)^k,\qquad
 C(d)=\sum_{h\in\mathbb Z}\gamma_h\gamma_{h+d}.
\]
Whenever the sums are absolutely convergent, Gaussian fourth-moment
factorization applied to
$u_t=\gamma_0^{-1}\sum_{k=0}^K b_kY_tY_{t-k}$ gives
\begin{equation}
 \Omega=\frac1{\gamma_0^2}\sum_{j=0}^K\sum_{k=0}^K
 b_jb_k\{C(k-j)+C(k+j)\}.
 \label{eq:gaussian-omega}
\end{equation}
Indeed,
\[
 \operatorname{Cov}(Y_0Y_{-j},Y_{-h}Y_{-h-k})
 =\gamma_h\gamma_{h+k-j}+\gamma_{h+k}\gamma_{h-j},
\]
and summing over $h$ yields the two convolutions in
\eqref{eq:gaussian-omega}. The expectation term cancels from the covariance.

For \eqref{eq:noise-family}, $\gamma_0=1$ and
$\gamma_h=ca^{|h|}+(1-c)\mathbf1_{\{h=0\}}$. For $0<a<1$,
\begin{equation}
 \begin{split}
 C(d)={}&c^2a^{|d|}
 \left(|d|+1+\frac{2a^2}{1-a^2}\right)
 +2c(1-c)a^{|d|}+(1-c)^2\mathbf1_{\{d=0\}}.
 \end{split}
 \label{eq:closed-convolution}
\end{equation}
The first term follows by separating $h$ between the two endpoints from
the two geometric tails; the other terms arise from the white-noise
contribution at zero lag.

An alternative positivity check is useful for stable Gaussian autoregressive
models. Under the convention
$\gamma_h=\int_{-\pi}^{\pi}e^{ih\omega}f(\omega)\,d\omega$,
Parseval's identity in \eqref{eq:gaussian-omega} gives
\begin{equation}
 \Omega=\frac{4\pi}{\gamma_0^2}
 \int_{-\pi}^{\pi}
 \left\{b_0+\sum_{k=1}^K b_k\cos(k\omega)\right\}^2
 f(\omega)^2\,d\omega.
 \label{eq:spectral-omega}
\end{equation}
For a stable Gaussian AR(2) with positive innovation variance,
$f(\omega)>0$. The trigonometric polynomial is not identically zero,
because its coefficient of $\cos(\omega)$ is $b_1=1$.
Thus $\Omega>0$. This argument establishes positivity, not numerical
uniqueness of a projection minimizer or its curvature sign.

\subsection{Exact counterexamples}
\label{app:exact-examples}
For \eqref{eq:noise-family} with $a=3/4$, $c=4/7$, and $K=2$,
$\rho_1=3/7$ and $\rho_2=9/28$. The half derivative of the objective is
\[
 \tfrac12 S_2'(p)=2p^3+\tfrac5{14}p-\tfrac37.
\]
Its derivative $6p^2+5/14$ is strictly positive for every real $p$.
The half derivative vanishes at $p=1/2$, so this is the unique global minimizer whenever
the parameter interval contains it in its interior.
The residuals are $R_1=-1/14$ and $R_2=1/14$, giving $q=2$ and $A=13/7$.
Substitution of $b_0=-3/4$, $b_1=b_2=1$ in
\eqref{eq:gaussian-omega}--\eqref{eq:closed-convolution} gives
$\Omega=52963/10976>0$.
The process is a measurable function of a stationary Gaussian AR(1) state
and independent Gaussian white noise. It has geometric mixing, moments
of every order, and the finite-dimensional moment CLT used above.
No boundary or zero-timescale singularity is involved.
For the correctly specified AR(1) control with $a=1/2$, $c=1$, $K=2$,
the corresponding values are $p^\star=1/2$, $q=A=2$, and $\Omega=57/16$.

For \eqref{eq:ma-example}, $0<\theta<1$ and
$\theta/(1+\theta^2)=11/27=:r$. The process has unit variance and is
1-dependent. Its objective is
\[
 S_2(p)=(r-p)^2+p^4,\qquad
 \tfrac12S_2'(p)=p+2p^3-r,\qquad
 \tfrac12S_2''(p)=1+6p^2>0.
\]
Since $r=1/3+2/27$, the unique global minimum is $p^\star=1/3$,
with $q=13/9$ and $A=5/3$.
Here
\[
 C(0)=1+2r^2=\frac{971}{729},\quad C(1)=2r=\frac{22}{27},
 \quad C(2)=r^2=\frac{121}{729},\quad C(d)=0\quad(|d|\geq3).
\]
Using $b_0=-11/27$, $b_1=1$, and $b_2=2/3$ in
\eqref{eq:gaussian-omega} yields
$\Omega=1121518/531441>0$.
Finite dependence and Gaussian moments verify the probabilistic input.
Both counterexamples therefore satisfy regularity independently of a
finite-sample optimizer or a HAC bandwidth choice.

\subsection{Direction within the observation-noise family}
\label{app:direction-family}
Take $\mathcal P=[-r,r]$ with $a<r<1$ in
\eqref{eq:noise-family}. Replacing a negative $p$ by $|p|$ strictly
reduces the odd-lag squared residuals and leaves the even-lag terms
unchanged, so a global minimizer is nonnegative.
Moreover, $G(0;\rho)=ca>0$, and $G(p;\rho)<0$ for $p\geq a$.
A regular interior minimizer must therefore satisfy $0<p^\star<a$.
The sequence
\[
 h_k=\frac{R_k}{(p^\star)^k}=c(a/p^\star)^k-1
\]
is strictly increasing. Since the first-order condition can be written
$\sum_k k(p^\star)^{2k-1}h_k=0$ with positive weights, its nonzero
entries have the negative-to-positive sign pattern. The strict conclusion
$A<q$ follows from Proposition~\ref{prop:single-crossing}.
This argument does not claim uniqueness for every parameter choice;
regularity is retained as a condition.
For $a<0$, simultaneous replacement of $(a,p)$ by $(-a,-p)$
preserves the objective and both curvatures on a symmetric parameter
interval. Changing the sign of $a$ therefore does not by itself reverse
the SE ratio within this family.

\subsection{Population Bartlett and finite-window expectations}
\label{app:quadratic-bartlett}
Let $W_t=(Y_t,Y_{t-1},\ldots,Y_{t-K})^\top$ and define a symmetric matrix
$H$ by
\[
 H_{0k}=H_{k0}=b_k/2,\quad
 H_{kk}=-b_k(p^\star)^k\quad(1\leq k\leq K),\qquad H_{00}=0,
\]
with all other entries zero. Then $v_t=\gamma_0^{-1}W_t^\top HW_t$.
If $(\Gamma_h)_{ij}=\gamma_{h+j-i}$, Gaussian quadratic-form covariance gives
\begin{equation}
 c_v(h)=\frac2{\gamma_0^2}
 \operatorname{tr}(H\Gamma_hH\Gamma_h^\top).
 \label{eq:quadratic-covariance}
\end{equation}
It can also be checked by expanding each product
$b_kY_tY_{t-k}-b_k(p^\star)^kY_{t-k}^2$ and using the
fourth-moment identity term by term.

For example, the lagged-square score in the MA(1) counterexample has
\[
 H=\begin{pmatrix}
 0&1/2&1/3\\
 1/2&-1/3&0\\
 1/3&0&-2/27
 \end{pmatrix}
\]
and exact covariances
\[
 c_v(0)=\frac{101120}{59049},\qquad
 c_v(1)=\frac{133439}{531441},\qquad
 c_v(2)=-\frac{3080}{59049},\qquad c_v(h)=0\quad(h\geq3).
\]
Their bilateral sum equals $1121518/531441$, agreeing with the
current-square calculation despite different individual covariances.

Let $V_M^{\,0}$ denote \eqref{eq:nw-v} calculated from the population
score rather than a fitted score. Because $\mathbb E v_t=0$,
\begin{equation}
 n\,\mathbb E V_M^{\,0}
 =c_v(0)+2\sum_{h=1}^M
 \left(1-\frac{h}{M+1}\right)\left(1-\frac hn\right)c_v(h).
 \label{eq:finite-window-bartlett}
\end{equation}
This finite-window identity is exact for known $p^\star$, $\mu$, and
$\gamma_0$. It is not an exact expectation formula for the fitted,
centered, standardized score used in a finite sample.
The limit $\Omega_M$ in Proposition~\ref{prop:fixed-m} is nevertheless
the same under its assumptions. This distinction separates an asymptotic
fixed-bandwidth discrepancy from additional plug-in and finite-sample effects.

\subsection{Equality of the curvatures without an exact fit}
\label{app:equality}
The sufficiency-versus-necessity distinction following
Theorem~\ref{thm:asymptotic} can be illustrated by a valid ACF.
Start with a unit-variance AR(1) spectrum with coefficient $1/2$,
and add
\[
 \frac{t}{\pi}\left\{\frac34\cos\omega-\frac32\cos(2\omega)
                    +\cos(3\omega)\right\},\qquad t=\frac1{100}.
\]
The original spectral density is bounded below by $1/(6\pi)$,
whereas the perturbation has absolute value at most $13t/(4\pi)$.
The resulting spectral density is strictly positive and integrates to one,
so it defines a stationary Gaussian process. Its first three
autocorrelations are $(1/2,1/4,1/8)+t(3/4,-3/2,1)$, and the remaining
autocorrelations are unchanged.
For $K=3$ and $\mathcal P$ containing $1/2$ in its interior,
\[
 S_3(p)-S_3(1/2)
 =(p-1/2)^2\{1+(p+1/2)^2+(p^2+p/2+1/4)^2-2t(p-1/2)\}.
\]
The expression in braces is positive for $p\in(-1,1)$.
Thus $p^\star=1/2$ is the unique minimizer, but its residual vector
$t(3/4,-3/2,1)$ is nonzero and orthogonal to both $m'(p^\star)$ and
$m''(p^\star)$. Consequently $A=q=41/16$.
This construction demonstrates that nonzero projection residuals are compatible with $A=q$.

\section{Complete numerical reporting and computation}\label{app:numerical}
The following tables present all specified A2 bandwidth results for the two feasible sample-bread procedures, the full primary-cell factorial comparison, and both population-full oracle conventions. The machine-readable summaries additionally retain all eight arms, paired contrasts, and interactions in every cell. Oracle arms repeat across $M$ because they do not use an estimated HAC bandwidth. Pilot results were analyzed separately and were not pooled with A2.

\begin{table}[htbp]\centering\small
\caption{Primary-cell factorial coverage, $K=36,T=3600,M=5$. Every row uses the same fitted parameter and 4,000 series. Population quantities are diagnostic substitutions.}\label{tab:factorial}
\begin{tabular}{llrr}\toprule
Bread & Score-mean variance & $p$ coverage (\%) & $\tau$ coverage (\%)\\\midrule
Sample $q$ & HAC & 77.825 & 78.200 \\
Sample $q$ & $\Omega/n$ & 87.375 & 87.000 \\
Population $q$ & HAC & 79.325 & 80.200 \\
Population $q$ & $\Omega/n$ & 86.725 & 85.875 \\
Sample $A$ & HAC & 90.500 & 89.675 \\
Sample $A$ & $\Omega/n$ & 94.900 & 95.400 \\
Population $A$ & HAC & 89.100 & 87.975 \\
Population $A$ & $\Omega/n$ & 95.275 & 94.850 \\
\bottomrule\end{tabular}

\end{table}

\begin{table}[htbp]\centering\small
\caption{Population-full oracle coverage under the $n=T-K$ and $T$ conventions (percent). Cell abbreviations are defined in \cref{tab:all-coverage}. Neither oracle is an exact finite-sample sampling variance.}\label{tab:oracle-conventions}
\begin{tabular}{lrrrr}\toprule
Cell & $p$: $\Omega/n$ & $p$: $\Omega/T$ & $\tau$: $\Omega/n$ & $\tau$: $\Omega/T$\\\midrule
C05-3600 & 94.925 & 94.750 & 94.850 & 94.825 \\
C075-3600 & 94.575 & 94.475 & 93.950 & 93.850 \\
N075-3600 & 95.275 & 95.225 & 94.850 & 94.825 \\
N09-3600 & 94.675 & 94.500 & 91.850 & 91.775 \\
C05-1000 & 96.500 & 96.000 & 95.350 & 95.100 \\
C05-10000 & 95.850 & 95.850 & 95.450 & 95.400 \\
N075-1000 & 95.150 & 94.650 & 92.850 & 92.700 \\
N075-10000 & 93.900 & 93.850 & 93.950 & 93.850 \\
\bottomrule\end{tabular}

\end{table}

\begin{table}[htbp]\centering\small
\caption{Empirical SD versus population scales using the original A2 $n$ convention. Compare with \cref{tab:sd}, which uses $T$.}\label{tab:sd-n}
\begin{tabular}{lrrrr}
\toprule
Process and length & $s_{A,n}$ & $s_{q,n}$ & $s/s_{A,n}$ & $s/s_{q,n}$ \\
\midrule
AR1 $(.5)$, $T=3600$ & 0.01973 & 0.01973 & 1.002 & 1.002 \\
AR1 $(.75)$, $T=3600$ & 0.01801 & 0.01801 & 1.004 & 1.004 \\
AR1 $(.75)$+noise, $T=3600$ & 0.02789 & 0.02081 & 0.985 & 1.320 \\
AR1 $(.9)$+noise, $T=3600$ & 0.02426 & 0.01673 & 0.996 & 1.444 \\
AR1 $(.5)$, $T=1000$ & 0.03794 & 0.03794 & 0.955 & 0.955 \\
AR1 $(.5)$, $T=10000$ & 0.01180 & 0.01180 & 0.990 & 0.990 \\
AR1 $(.75)$+noise, $T=1000$ & 0.05363 & 0.04001 & 0.980 & 1.314 \\
AR1 $(.75)$+noise, $T=10000$ & 0.01668 & 0.01245 & 1.011 & 1.356 \\
\bottomrule
\end{tabular}

\end{table}
\FloatBarrier
\clearpage
\begingroup\footnotesize\setlength{\tabcolsep}{5pt}
\begin{longtable}{llrrrrr}
\caption{All A2 bandwidth results. Rates are percentages. C05 and C075 are exact AR1 controls; N075 and N09 are noise mixtures with $c=4/7$. The number after the model code is $T$. Each $T=3600$ cell has $R=4000$; other cells have $R=2000$. All rows within a cell share the same series.}\label{tab:all-coverage}\\[6pt]
\toprule
Cell & $M$ & GN $p$ & Full $p$ & GN $\tau$ & Full $\tau$ & $\Delta p$ [MC 95\%] \\
\midrule\endfirsthead
\multicolumn{7}{c}{\tablename\ \thetable\ (continued)}\\
\toprule Cell & $M$ & GN $p$ & Full $p$ & GN $\tau$ & Full $\tau$ & $\Delta p$ [MC 95\%] \\\midrule\endhead
\bottomrule\endfoot
C05-3600 & 5 & 92.825 & 93.175 & 93.200 & 92.850 & 0.350 [-0.010, 0.710] \\
C05-3600 & 8 & 93.550 & 93.675 & 93.600 & 93.475 & 0.125 [-0.211, 0.461] \\
C05-3600 & 16 & 94.025 & 94.050 & 94.025 & 93.825 & 0.025 [-0.318, 0.368] \\
C05-3600 & 32 & 94.175 & 94.050 & 94.225 & 94.000 & -0.125 [-0.482, 0.232] \\
C05-3600 & 64 & 94.075 & 93.675 & 94.025 & 93.925 & -0.400 [-0.804, 0.004] \\
\addlinespace
C075-3600 & 5 & 88.225 & 87.800 & 87.700 & 87.600 & -0.425 [-0.927, 0.077] \\
C075-3600 & 8 & 90.725 & 90.375 & 90.100 & 89.650 & -0.350 [-0.850, 0.150] \\
C075-3600 & 16 & 92.400 & 92.000 & 91.675 & 91.425 & -0.400 [-0.865, 0.065] \\
C075-3600 & 32 & 93.125 & 92.675 & 92.325 & 92.150 & -0.450 [-0.888, -0.012] \\
C075-3600 & 64 & 93.225 & 93.025 & 92.675 & 92.025 & -0.200 [-0.627, 0.227] \\
\addlinespace
N075-3600 & 5 & 77.825 & 90.500 & 78.200 & 89.675 & 12.675 [11.644, 13.706] \\
N075-3600 & 8 & 80.700 & 91.825 & 81.050 & 91.375 & 11.125 [10.150, 12.100] \\
N075-3600 & 16 & 83.375 & 93.050 & 83.375 & 92.875 & 9.675 [8.759, 10.591] \\
N075-3600 & 32 & 84.750 & 93.650 & 83.975 & 93.400 & 8.900 [8.017, 9.783] \\
N075-3600 & 64 & 84.725 & 93.625 & 84.075 & 93.400 & 8.900 [8.017, 9.783] \\
\addlinespace
N09-3600 & 5 & 60.175 & 77.775 & 59.350 & 76.900 & 17.600 [16.420, 18.780] \\
N09-3600 & 8 & 65.850 & 82.725 & 65.475 & 82.125 & 16.875 [15.714, 18.036] \\
N09-3600 & 16 & 73.075 & 88.850 & 72.575 & 86.900 & 15.775 [14.645, 16.905] \\
N09-3600 & 32 & 77.075 & 91.575 & 76.600 & 89.625 & 14.500 [13.409, 15.591] \\
N09-3600 & 64 & 78.275 & 92.250 & 78.475 & 90.425 & 13.975 [12.900, 15.050] \\
\addlinespace
C05-1000 & 5 & 93.750 & 92.300 & 93.000 & 91.600 & -1.450 [-2.174, -0.726] \\
C05-1000 & 8 & 93.900 & 93.000 & 93.300 & 91.850 & -0.900 [-1.578, -0.222] \\
C05-1000 & 16 & 93.850 & 93.050 & 93.550 & 92.200 & -0.800 [-1.546, -0.054] \\
C05-1000 & 32 & 93.700 & 92.900 & 93.200 & 91.850 & -0.800 [-1.533, -0.067] \\
C05-1000 & 64 & 92.650 & 91.450 & 92.300 & 90.950 & -1.200 [-1.891, -0.509] \\
\addlinespace
C05-10000 & 5 & 93.900 & 94.300 & 94.150 & 94.100 & 0.400 [0.091, 0.709] \\
C05-10000 & 8 & 94.650 & 94.550 & 94.650 & 94.450 & -0.100 [-0.410, 0.210] \\
C05-10000 & 16 & 95.100 & 94.950 & 94.750 & 94.800 & -0.150 [-0.503, 0.203] \\
C05-10000 & 32 & 95.300 & 95.000 & 95.000 & 95.250 & -0.300 [-0.667, 0.067] \\
C05-10000 & 64 & 95.000 & 95.250 & 95.150 & 95.250 & 0.250 [-0.103, 0.603] \\
\addlinespace
N075-1000 & 5 & 77.500 & 88.850 & 76.550 & 87.150 & 11.350 [9.959, 12.741] \\
N075-1000 & 8 & 79.500 & 90.350 & 78.950 & 88.500 & 10.850 [9.487, 12.213] \\
N075-1000 & 16 & 81.100 & 91.500 & 80.900 & 89.600 & 10.400 [9.062, 11.738] \\
N075-1000 & 32 & 82.000 & 91.250 & 80.950 & 89.700 & 9.250 [7.980, 10.520] \\
N075-1000 & 64 & 81.450 & 90.250 & 80.150 & 89.300 & 8.800 [7.558, 10.042] \\
\addlinespace
N075-10000 & 5 & 77.950 & 89.250 & 77.600 & 89.150 & 11.300 [9.912, 12.688] \\
N075-10000 & 8 & 80.550 & 91.050 & 80.100 & 91.000 & 10.500 [9.156, 11.844] \\
N075-10000 & 16 & 82.800 & 92.350 & 82.550 & 92.400 & 9.550 [8.262, 10.838] \\
N075-10000 & 32 & 83.750 & 93.050 & 84.100 & 92.900 & 9.300 [8.027, 10.573] \\
N075-10000 & 64 & 84.500 & 92.950 & 84.100 & 93.100 & 8.450 [7.231, 9.669] \\
\addlinespace
\end{longtable}
\endgroup

\subsection{Monte Carlo uncertainty of an empirical standard deviation}
For independent replicate estimates $P_1,\ldots,P_R$, let $s$ be their empirical SD and $v_r=(P_r-\bar P)^2$. The post hoc SD error bars use the delta approximation
\begin{equation}
 \widehat{\operatorname{MCSE}}(s)
 =\frac{\operatorname{sd}(v_1,\ldots,v_R)}{2s\sqrt R}.
\end{equation}
For a ratio to a known population scale, this MCSE is divided by that scale. This large-$R$ approximation uses the empirical fourth moment. The normal-theory relative MCSE $[2(R-1)]^{-1/2}$ provides a reference value; the displayed error bars use the empirical-moment calculation.

\subsection{Deterministic source-design calculations}
Population targets were computed from the stated ACF and the same fixed-$K$ objective. Numerical validation combined enumeration of real stationary roots and endpoints, a dense grid, high-precision root refinement, and finite-difference curvature estimates. These calculations apply to the evaluated parameter grid.

For the source AR2 specification, the ACF starts at $\rho_0=1$, $\rho_1=0.8$ and satisfies $\rho_k=\phi_1\rho_{k-1}+\phi_2\rho_{k-2}$. The printed radii $0.5,0.642,0.758,0.859,0.949$ were treated as exact. Residual sign-change counts were computed to assess the applicability of the single-crossing condition.

\subsection{Reference implementation}\label{app:reference-code}
The comparison uses the public AD implementation at commit
\begin{quote}\small\ttfamily
cd0072ee3b42f271e49de3682d5f7a41bb9070d2
\end{quote}
of the \texttt{griegner/fmri-timescales} repository (accessed 16 November 2026). At this commit, \href{https://github.com/griegner/fmri-timescales/blob/cd0072ee3b42f271e49de3682d5f7a41bb9070d2/fmri_timescales/acf_utils.py}{\texttt{acf\_utils.py}} centers and standardizes each series with a variance divisor of $T$, computes lag products by convolution, and divides each lag sum by $T$. For finite, nonconstant inputs, this gives the biased ACF in \eqref{eq:sample-projection}. The AD routine in \href{https://github.com/griegner/fmri-timescales/blob/cd0072ee3b42f271e49de3682d5f7a41bb9070d2/fmri_timescales/timescale_utils.py}{\texttt{timescale\_utils.py}} uses the lagged-square score and normalization described in Section~\ref{sec:methods}. The original article's execution commit is unknown.

\end{document}